\documentclass[12pt,a4paper,twocolumn]{narms}

\usepackage{graphicx}
\usepackage{psfrag}

\usepackage{subfigure} 
\usepackage{psfig}
\usepackage{times}   
\usepackage{chicnarm}
\usepackage{topcapt}

\makeatletter
\renewcommand{\section}{\@startsection%
{section}%
{1}%
{0mm}%
{- \baselineskip}%
{0.15\baselineskip}%
{\normalfont\normalsize}}%

\renewcommand{\subsection}{\@startsection
{subsection}%
{2}%
{0mm}%
{-\baselineskip}%
{0.15\baselineskip}%
{\normalfont\normalsize}}%
\makeatother

\begin{document}

\title{Discrete numerical simulation, quasistatic deformation and the origins of strain
in granular materials.}
\author{
\large {Ga\"el Combe \&  Jean-No\"el Roux}\\
{\em Laboratoire des Mat\'eriaux et des Structures
du G\'enie Civil,
Institut Navier, Champs-sur-Marne, France}\\
}
\date{}

\abstract{{\bf ABSTRACT:}
Systematic numerical simulations of model dense granular materials in monotonous,
quasistatic deformation reveal the existence of two different r\'egimes. In the
first one, the macroscopic strains stem from the deformation of contacts. The 
motion can be calculated by purely static means, without inertia,
stress controlled or strain rate controlled simulations yield identical
smooth rheological curves for a same sample. In the second r\'egime, strains are
essentially due to instabilities of the contact network, the approach to the limits
of large samples and of small strain rates is considerably slower and the material is
more sensitive to perturbations. These results are discussed and related to
experiments~: measurements of elastic moduli with very small strain increments, and
slow deformation (creep) under constant stress.
}


\maketitle
\frenchspacing   


\section{INTRODUCTION}

Despite its now widespread use~\shortcite{KI01}, discrete numerical simulation of granular materials, motivated either by
the investigation of small scale (close to the grain size) phenomena, or by the study of
microscopic origins of
known macroscopic laws, still faces difficulties. Microscopic parameters, some of which
are to be defined at the (even smaller)
scale of the contact, are incompletely known. Macroscopic constitutive laws do not emerge easily out of noisy
simulation curves, and the numerically observed \emph{dynamic} sequences of rearrangements
might appear to contradict the traditional
macroscopic \emph{quasistatic} assumption.
Detailed and quantitative comparisons with experiments can be used to adjust microscopic models, but 
a systematic exploration of the effect of the various parameters
throughout some admissible range is also worthwhile. This is the purpose of the present study, 
which also addresses the fundamental issues of the macroscopic and quasistatic limits, in the case of
the biaxial compression of dense, two-dimensional (2D) samples of disks.

In section 2, we introduce the model and the numerical methods and define dimensionless
parameters that are robust indicators of the relative importance
of different phenomena. Rheological curves can be evaluated
in the large sample limit (section 3), and
their sensitivity to parameters assessed.
We observe (section 4) two different mechanical r\'egimes, according to whether
the dominant microscopic origin of strain is material deformation in the contacts or
rearrangements of the contact network. Connections to
some experimental observations are suggested in part 5, while the conclusion section
outlines further perspectives.
\section{NUMERICAL MODEL AND PROCEDURES}
\subsection{\em Grain-level mechanics \label{subsec:micro}}
Our computational procedure is one of the simplest types of `molecular dynamics' or
`discrete element' method~\shortcite{CUND79} for solid grains.
We consider 2D assemblies of disks, with diameters uniformly distributed between
$a/2$ and $a$, and masses and moments of inertia evaluated accordingly (as for
homogeneous solid cylinders of equal lengths).
$m$ will denote the mass of a disk of diameter $a$, and $N$ the number of disks.

These grains
interact in their contacts with a linear elastic law and Coulomb
friction. The normal contact force $F_N$ is thus related to the normal deflection (or
apparent interpenetration) $h$ of the contact as $F_N = K_N h Y(h)$, $Y$ being the
Heaviside step function
(equal to $1$ for $h>0$, to $0$ otherwise).
The tangential component $F_T$ of the contact force is proportional to the
tangential elastic relative displacement, with a tangential stiffness coefficient $K_T$. The
Coulomb condition $\vert F_T\vert \le \mu F_N$ requires an incremental evaluation of 
 $F_T$ every time step, which leads to some amount of slip each time one of the
equalities $F_T=\pm \mu F_N$ is imposed.
A normal viscous component 
opposing the relative normal motion of any pair of grains in contact is also added
to the elastic force $F_N$. Such a term -- of unclear physical origin
in dense multicontact systems -- is often introduced
to ease the approach to mechanical equilibrium. Its influence will be assessed in part 3.
The viscous force is 
proportional to the normal relative velocity, and the damping coefficient in the contact 
between grains $i$ and $j$ is a constant fraction $\zeta$ ($0\le \zeta \le 1$) of the critical value
$2({K_Nm_im_j\over m_i+m_j})^{1/2}$. 
(In a binary collision the normal `restitution coefficient'
is $0$ for $\zeta=1$ and $1$ for $\zeta=0$). $\zeta$,  $K_N$, $K_T$, and $\mu$ are the same in all contacts.
The motion of grains is calculated on solving Newton's equations.
\subsection{\em Numerical compression tests}
Two different types of boundary conditions
are used~: either the container walls are physical objects, with masses, satisfying
Newton's equations (but requested to move in the direction perpendicular to their
orientation), or periodic boundary conditions (no walls) are implemented. In both
cases, the changes in cell size and shape under controlled stress
involves specific dynamical parameters which could be discussed in more
detail. Here we will simply deem such parameter choice innocuous if results are reproducible, size-independent
and consistent.
We use soil mechanics sign conventions for stresses and strains.
Samples are first compressed isotropically under a constant
pressure $P$.
Once a mechanical equilibrium is reached under pressure $P$, samples are submitted to biaxial compression tests.
The lateral stress,
$\sigma_1$ is maintained equal to $P$, while either $\epsilon_2$ is increased at a constant rate
$\dot \epsilon_2$  (a procedure hereafter referred to as SRC, for \emph{strain rate controlled})
or $\sigma_2$ is stepwise increased by small fractions of $P$ , and one waits for
the next equilibrium configuration before changing $\sigma_2$ (a SIC,
for \emph{stress increment controlled}, procedure). In the sequel $q$ denotes the ratio
$(\sigma_2-\sigma_1)/\sigma_1$, while $\epsilon_2$ and
$\epsilon _v = \epsilon _1 + \epsilon _2 - \epsilon _1\epsilon _2$
are respectively termed `axial' and `volumetric' strain, in analogy with 3D
axisymmetrical triaxial tests. 
\subsection{\em Dimensional analysis}
Rheological curves and internal sample states obtained in monotonous biaxial tests are defined in the macroscopic limit
$N\to \infty$. If expressed by relations between
dimensionless quantities $\epsilon _2$, $q$, $\epsilon_v$, they should depend on the
friction coefficient $\mu$ and on ratio $K_T/K_N$, and on three other dimensionless
parameters: 
$\kappa = K_N / P$, the \emph{stiffness parameter}, which expresses
the level of contact deformation, 
$\gamma = \dot \epsilon _2 \sqrt{m/P}$, the \emph{inertia parameter}, evaluating, in SRC
(constant $\dot \epsilon _2$) tests, the importance of dynamical effects, and
$\zeta$, the \emph{damping parameter}, introduced in paragraph~\ref{subsec:micro},
characterizing viscous dissipation. 
 The contact coordination number is a decreasing function of $\kappa$.
The quasistatic limit is the limit of small $\gamma$.
\section{BIAXIAL COMPRESSION OF DENSE SYSTEMS~: RESULTS}
\subsection{\em Preparation, initial states, procedures.}
The sample preparation procedure is well known to exert a strong influence on the
mechanical properties of a granular sample as, in particular, dense or loose initial states respond
differently~\shortcite{Muirwood} to
load increments. Moreover, experiments also showed that density is not sufficient to determine
the behaviour in a triaxial test~\shortcite{BENA1999}. Numerical simulations may in principle attempt to imitate as closely
as possible laboratory experiments. The simulations of such processes as deposition under gravity within a walled container
is however difficult, as it requires 
large number of particles. Inhomogeneous states one obtains in such cases request
samples much larger then a representative volume element, which is itself 
much larger than the grain size. 
Moreover, the transition from an initial fluid-like configuration to a solid-like grain assembly 
is bound to be sensitive to static and dynamic parameters~\shortcite{SEGHL01}.

Here we focus on the slow quasistatic deformation of certain types of granular assemblies, once they have been
prepared in some well defined initial state. Therefore we leave a detailed (and necessary)
study of the preparation process to future research, and adopt a simple numerical procedure which provides us
with homogeneous, reproducible, sample size -independent initial states in equilibrium under an isotropic pressure.
The numerical procedure is an isotropic, monotonous compaction from an initial gas-like
configuration with a solid fraction $\Phi$ 
of about $20\%$. To obtain a dense sample, a different, smaller value is attributed to the coefficient
of friction in this initial dynamic compression step. 
Two series of samples are studied here. The first one -- called series A hereafter -- was prepared between solid,
frictionless walls. It was observed in that case that one had to set $\mu$ to zero in the preparation stage if we were
to obtain a homogeneous stress field. Simulations of series A were therefore performed starting from the very dense
states which result from a compression without intergranular friction~\shortcite{COMB2001}.
The results below, some of which were presented
in~\shortcite{RC02}, were obtained with $\mu=0.25$ during biaxial compressions,
and a rigidity level $\kappa=10^5$. $K_T/K_N$ was
set to $1/2$. Biaxial tests were SIC, with small $q$ steps
$\delta q=10^{-3}$. Each successive mechanical equilibrium is deemed attained when the total force
(or torque) on each grain is less than $10^{-4}aP$ (resp. $10^{-4}a^2P$) and when the relative difference
between the internal overall stresses (deduced
from non-viscous intergranular forces)  and their prescribed values is less than $10^{-4}$. $\zeta$ was set to high values
(near $1$) and $N$ ranged from 1024 to 4900. In the initial isotropic state, the solid fraction
(extrapolated to $N\to \infty$)
is $\Phi=0.844\pm 0.001$, all but $5.5\%$ of the disks carry forces and the coordination number,
ignoring those inactive grains,
is $z\simeq 4.01$, very close to the isostatic limit~\shortcite{JNR2000} of $4$ reached with
rigid, frictionless disks in equilibrium.

\begin{figure}
 \centering 
 \includegraphics[width=9cm]{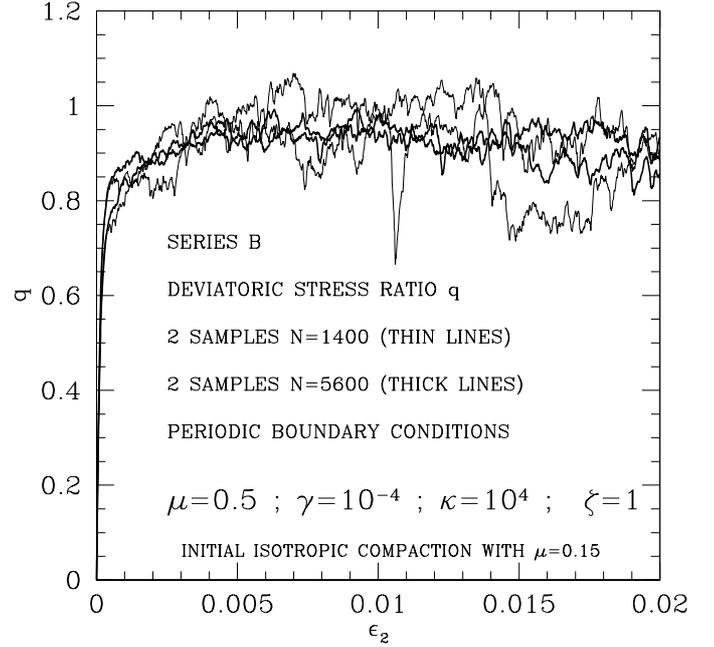}
\caption{
{\small
$q$ versus axial strain $\epsilon_2$ in B samples of 2 different sizes. Fluctuations are larger for the smaller samples.}
\label{fig:comptailleq}
}
\end{figure}

For the second series of simulations, series B, we used
periodic boundary conditions. Samples are thus devoid of edge effects. They shrink homogeneously in the isotropic
compression stage. Series B samples
were compressed with $\mu=0.15$, and subsequent biaxial tests performed with $\mu=0.5$.
Different stiffness levels, ($\kappa=10^3$, $10^4$ and $10^5$) were used, with $K_T/K_N$ fixed to $1$,
as well as different inertia
parameters $\gamma$ ($10^{-3}$, $10^{-4}$, sometimes $10^{-5}$). SRC tests were compared to SIC ones
(with $\delta q=10^{-2}$ and $\zeta \simeq 1$). Samples of 1400 and 5600 disks were simulated.
The initial solid fraction, due
to the finite $\mu$ value during compression, is lower than for A samples, as well as the coordination number $z$
among force-carrying disks. Values of $\Phi$, $z$, and the fraction of inactive disks $x_0$, for the investigated $\kappa$ 
values are given in table~\ref{tab:initB}.
\begin{table}
\topcaption{{\small Initial state data for series B simulations.}
\label{tab:initB}
}
 \begin{center}
 \begin{tabular}{|l||c|c|c|}  \cline{1-4}
  $\kappa$ & $\Phi$    & $z$    & $x_0$ (\%)     \\
\hline
\hline
$10^5\strut$   &$0.8226\pm 8.10^{-4}$ &$3.59\pm 2.10^{-2}$ & $10.0\pm 0.5$\\
\hline
$10^4\strut$   &$0.8230\pm 8.10^{-4}$ &$3.64\pm 2.10^{-2}$  & $9.0\pm 0.6$\\
\hline
$10^3\strut$   &$0.8258\pm 9.10^{-4}$ &$3.77\pm 8.10^{-3}$  & $6.7\pm 0.3$\\
\hline
\end{tabular}
\end{center}
\vskip -0.4cm
\end{table}
The typical aspect of $q$ versus $\epsilon _2$ curves is illustrated on fig.~\ref{fig:comptailleq},
for series B samples with $\kappa=10^4$ and $\gamma=10^{-4}$. 
They are characteristic of very dense samples, as in~\shortcite{KUHN99}.
\subsection{\em Stress-strain curves and macroscopic limit.}
\begin{figure}[b]
 \centering 
 \includegraphics[width=8.5cm]{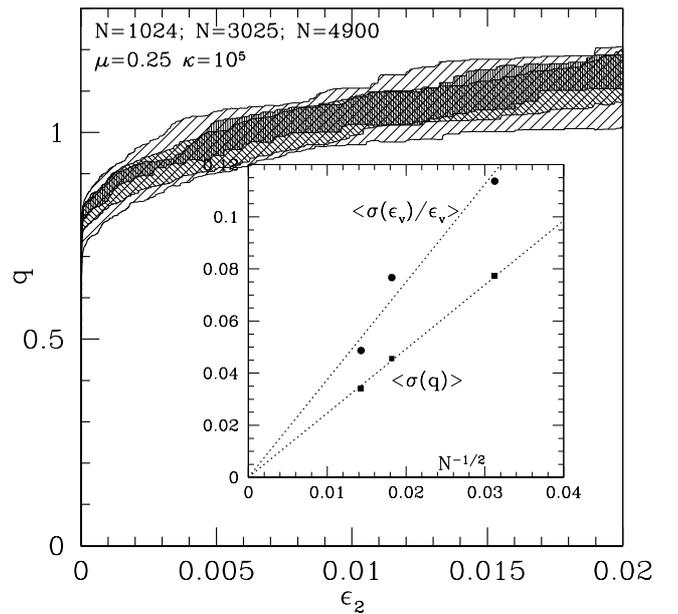}
\caption{
{\small
Hashed zone (the darker the larger $N$) one r.m.s. deviation on each side of average curve for the
3 sample sizes indicated (series A, SIC with $\delta q=10^{-3}$). Inset~: its average width over the $\epsilon _2\le 0.02$ interval,
versus $1/\sqrt{N}$, along with the average relative uncertainty on $\epsilon_v$}
\label{fig:courbstat}
}
\end{figure}

The increase of $q$ with $\epsilon_2$ is initially quite fast, $q$
reaching about $0.8$ for $\epsilon_2 < 10^{-3}$. Then the deviator stress keeps increasing and reaches
an apparent plateau for $\epsilon_2 \sim 0.01$. Those dense samples are markedly dilatant (fig.~\ref{fig:compvitv} below),
after
a very small initial contraction their
volume steadily increases, even after $q$ appears to have levelled off. The important stress
fluctuations in those SRC tests is striking on fig.~\ref{fig:comptailleq},
but are considerably reduced, as
well as sample-to-sample differences, as $N$ increases from $1400$ to $5600$. 
Dilatancy curves (see below)
are
smoother. 
Smooth stress-strain curves can thus be expected in
the macroscopic limit $N\to\infty$. This was more carefully
checked for simulation series $A$,
on studying three sample sizes~: on fig.~\ref{fig:courbstat}
the shaded zones extend to one standard deviation on each side of the
average curves, for
$q$ plotted as a function of $\epsilon _2$ for $N=1024$ (26 samples),  $N=3025$ (10 samples), and  $N=4900$ (7 samples). 
Fig.~\ref{fig:courbstat} does indicate a systematic decrease of the fluctuation level (see inset),
compatible with a regression as $N^{-1/2}$,
just like for an average over a number of independent contributions
(subsystems of representative size) proportional to $N$.
Series A samples respond in a similar way to deviator stresses as type B ones (although of course,
due to different
initial states, $\mu$ and $\kappa$, constitutive laws will differ). The initial increase of $q$,
so fast that it cannot
be distinguished from the axis on fig.~\ref{fig:courbstat}, is followed by a slower variation.
(Yet, unlike in the B case, $q$ does not reach a maximum for
$\epsilon _2\le 0.02$). 
`Volumetric' strains are also qualitatively similar for series A and B.
\subsection{\em Role of parameters $\zeta$, $\gamma$, $\kappa$.\label{subsec:role}}
\begin{figure}[b]
 \centering 
 \includegraphics[width=9cm]{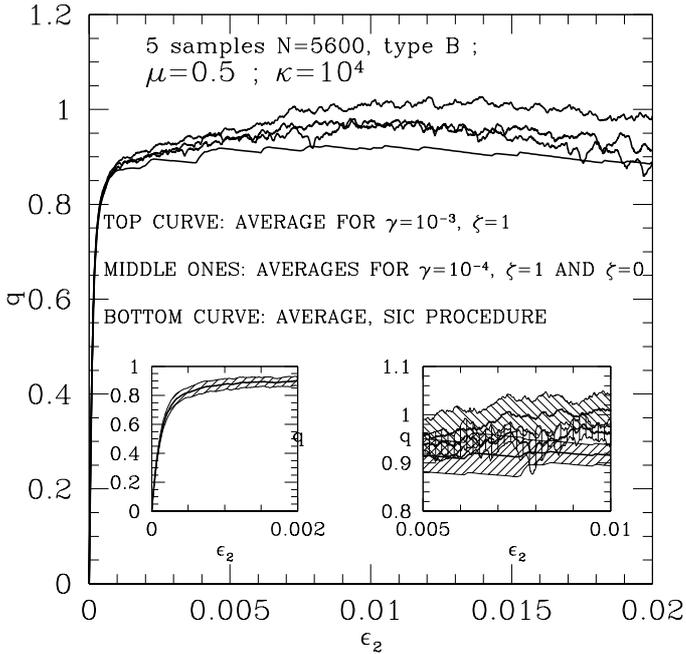}
\caption{
{\small
Average $q$ versus $\epsilon_2$ for conditions indicated. Left inset: detail of one curve with
r.m.s. deviations, small $\epsilon_2$.  Right inset: averages and r.m.s. deviations for $\gamma=10^{-3}$,
 $\gamma=10^{-4}$ and SIC tests.}
\label{fig:compvitq}
}
\end{figure}

\begin{figure}
\centering 
\includegraphics[width=9cm]{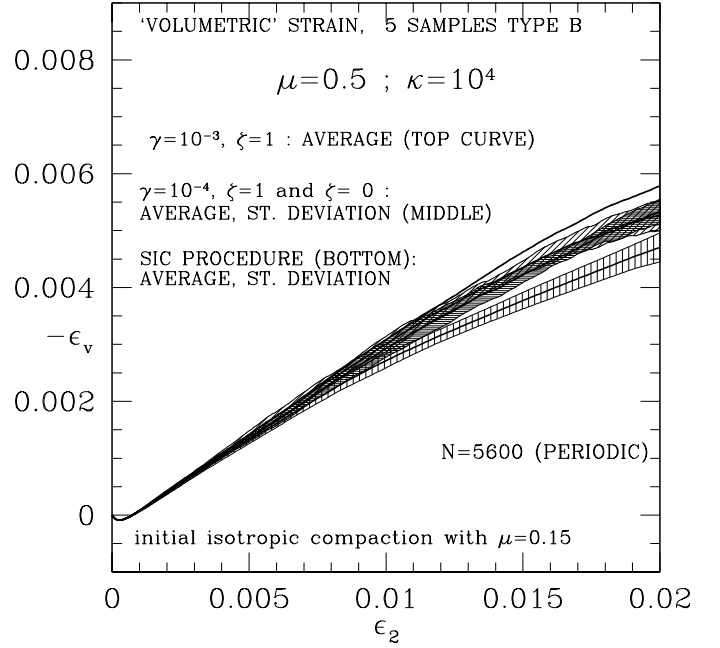}
\caption{
{\small
Same as fig.~\ref{fig:compvitq} for $\epsilon _v$ vs. $\epsilon _2$, standard deviations shown except for
uppermost ($\gamma=10^{-3}$) curve.}
\label{fig:compvitv}
}
\end{figure}

\begin{figure}[b]
\centering 
\includegraphics[width=9cm]{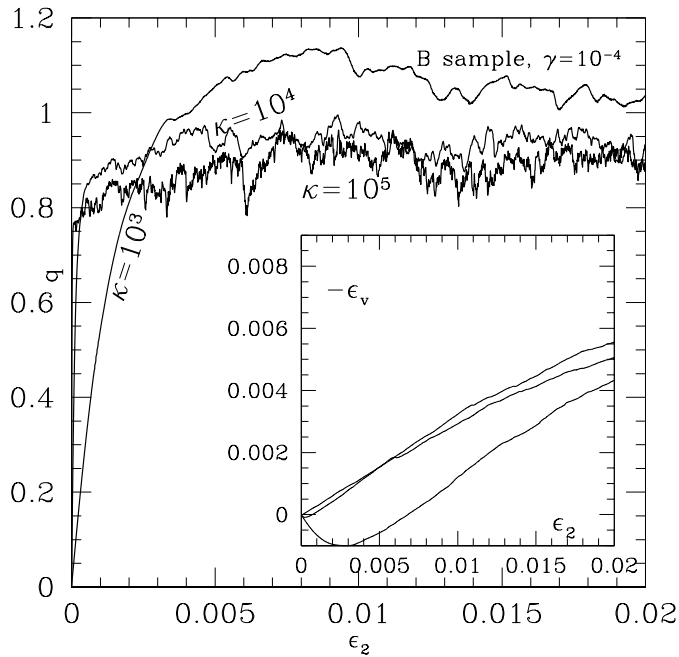}
\caption{
{\small
Results for one B-sample with 3 different stiffness values,
$q$ (main plot) and $\epsilon _v$ (inset) vs. $\epsilon_2$.}
\label{fig:compkappa}
}
\end{figure}

The quasistatic stress-strain  curve should be the same for SRC and SIC biaxial compressions, 
independent on $\zeta$ and on $\gamma$
if it is small enough. To check this, five samples of series B were submitted to SRC tests
with $\gamma=10^{-3}$ and $\zeta=1$, $\gamma=10^{-4}$ and $\zeta=1$,
$\gamma=10^{-4}$ and $\zeta=0$, and to SIC ones with $\delta q = 10^{-2}$.
Average curves for $q$ versus $\epsilon _2$
(fig.~\ref{fig:compvitq}) and $\epsilon _v$ versus $\epsilon _2$ (fig.~\ref{fig:compvitv})
for those 4 sets of simulations are displayed (and
standard deviations levels indicated as on fig.~\ref{fig:courbstat}).
Obviously, the value of $\zeta$ does not have any appreciable
influence on the rheological curve. Intergranular friction
is the dominating dissipation mechanism, and it can be checked
that the differences between stresses evaluated with and without
viscous forces differ by negligible amounts for all SRC tests.
However, results are affected by the reduced rate $\gamma$, or the choice of
an SIC procedure. A smaller $\gamma$  (according to  its definition, this amounts to
a slower compression, lighter grains or higher pressures) results in
 smaller deviator and 
dilatancy values for a given `axial' strain. SIC tests, as
one waits for equilibrium, are the slowest, and SIC curves
can be regarded as an extrapolation of SRC ones to $\gamma = 0$. (The occurrence of
slightly \emph{decreasing} $q$ values in SIC tests might seem surprising,
but is due to the use of real Cauchy stresses to draw the curve, while stresses
defined in terms of initial cell dimensions are used in the calculations). 
The effects of the \emph{stiffness parameter} $\kappa$ are illustrated on fig.~\ref{fig:compkappa}. It is most
apparent in the initial rise of $q$, which is the faster for higher $\kappa$, and the small-strain contractant
r\'egime (see inset), which develops with softer contacts.
For smaller $\kappa$, the packing appears indeed to be softer. The curves at
larger strains display no conspicuous difference between $\kappa=10^4$ and $\kappa=10^5$,
although the softest grains,
$\kappa=10^3$ appear to withstand a somewhat higher deviator stress. The dilatancy - slope of $-\epsilon _v$ versus
$\epsilon_2$ - is not affected. 
The time scale for stress fluctuation during monotonous tests at a given strain
rate is a strongly decreasing function of $\kappa$,
hence the smoother curves on fig.~\ref{fig:compkappa} for softer contacts. 
The effects of the parameters on rheological curves are related to some changes in the internal states
of the system undergoing compression. The effect of $\gamma$ is related to the greater distance to equilibrium of systems
under higher strain rate. Characteristic quantities are the average kinetic energy per particle, $e_c$ (in units of
$a^2P$) 
and the quadratic
average of the net force on a particle (in units of $aP$), $f_2$.
Those quantities tend to slowly increase with $\epsilon _2$ during the test, 
but typical values for $\epsilon _2 =0.01$ can be cited.
As for SIC tests, one only records equilibrium
positions, ensuring $f_2\le 10^{-5}$ and $e_c \le 10^{-8}$. The coordination number $z$
and the
proportion of sliding contacts $X_s$ vary quickly before $\epsilon _ 2=10^{-3}$
and remain essentially constant afterwards (one has $z=3.12$, on average, for $\kappa=10^4$ and $\gamma=10^{-4}$,
$z=3.05$ for $\kappa=10^4$ and $\gamma=10^{-3}$). 
Tests with the highest $\gamma$ values $10^{-3}$ are, logically, the farthest from equilibrium ($e_c=1.5\ 10^{-5}$, and
$f_2=0.01$, while $e_c\simeq 5\ 10^{-7}$ and $f_2=0.02$ for $\gamma =10^{-4}$). The change of $\kappa$ makes a
significantly larger difference from $10^4$ to $10^3$ than from $10^5$ to $10^4$. Unlike $e_c$ and $f_2$, which
essentially depend on $\gamma$, $z$ and $X_s$ are sensitive to both parameters.
In SIC tests ($\kappa=10^4$), $z$ decreases from its initial value to about $3.22$ (for $\epsilon_2 \sim 0.01$)
which is consistent with its dependence on $\gamma$ in SRC conditions.
Intermediate configurations of SIC tests, remarkably, do not have any sliding contact:
on approaching equilibrium, all
contact forces leave the edge of the Coulomb cone. Upon resuming an SRC motion,
very small displacements can mobilize friction and 
$X_s>0$ is observed (typically $X_s\simeq 10\%$, if $\gamma=10^{-4}$ and $\kappa=10^4$, $X_s$ increases with $\gamma$ and
with $\kappa$). 
\section{DIFFERENT ORIGINS OF STRAIN}
\begin{figure}
\centering 
\includegraphics[width=9cm]{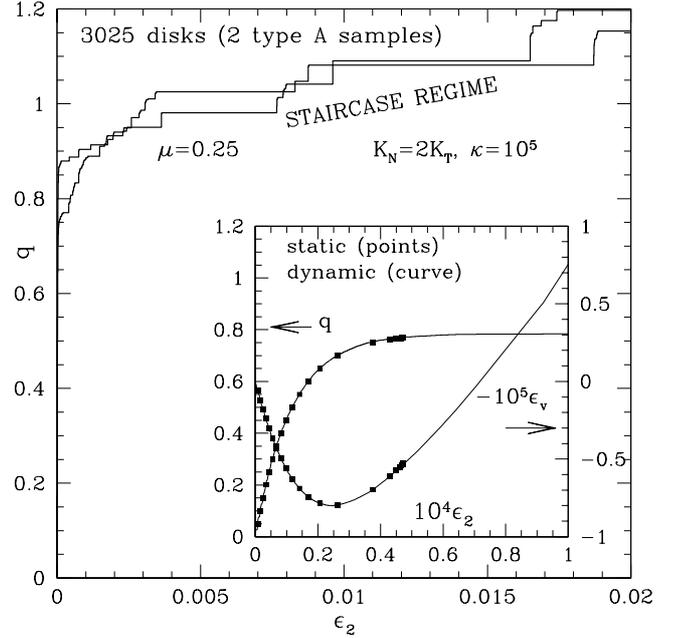}
\caption{
{\small
Two SIC $q$ vs. $\epsilon_2$ curves. Inset~:
initial strictly quasistatic r\'egime, blown-up $\epsilon$ scales. Results on one sample are identical with both static and
dynamic methods.}
\label{fig:dessmarfro}
}
\end{figure}

\begin{figure}
 \centering 
 \includegraphics[width=8.5cm]{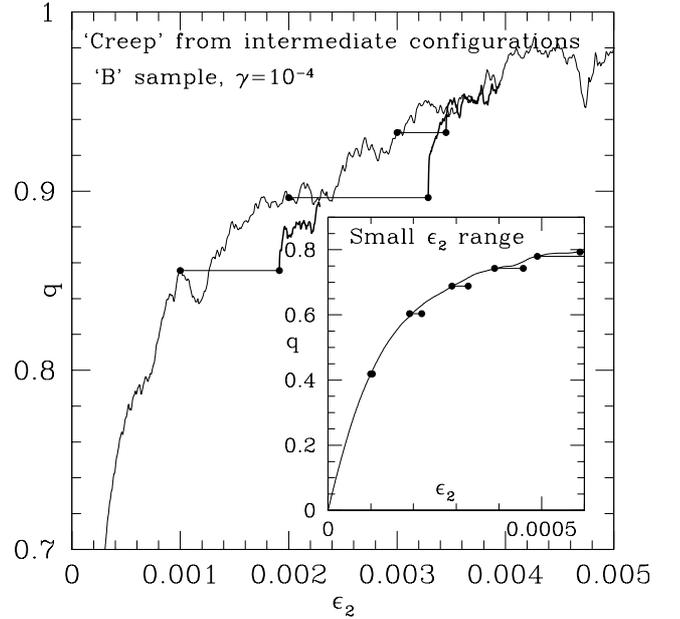}
\caption{
{\small
`Creep tests', dots on main plot showing initial and final (equilibrium) states. Effect of resuming 
compression SRC way shown as thick lines. Inset: creep tests within strictly quasistatic range.}
\label{fig:fluage}
}
\end{figure}

One striking aspect of the rheological curves is the existence of two different r\'egimes. At small $\epsilon_2$,
close to the initial isotropic state, 
curves are quite smooth and reproducible, sample
to sample fluctuations are very small (figs.~\ref{fig:comptailleq} and~\ref{fig:compvitq} ), SIC and SRC tests (whatever
$\gamma\le 10^{-3}$) are in
perfect agreement (figs.~\ref{fig:compvitq} and~\ref{fig:compvitv}), 
and $\kappa$  strongly affects the results (fig.~\ref{fig:compkappa}). Coordination numbers and friction mobilization
change fast from initial values
(table~\ref{tab:initB}) to the roughly constant ones given in paragraph~\ref{subsec:role}. 
At larger strains, the system is sensitive to the strain rate,
much more than to the stiffness parameter. Fluctuations are considerably larger, and the stepwise increase of $q$, as
one records the ensuing sequence of equilibria, results in a staircase-shaped $q$ versus $\epsilon_2$ curve, as on
fig.~\ref{fig:dessmarfro}. $q$ increments in those SIC simulations are very small, $\delta q=10^{-3}$, so that
nearly vertical segments on those plots correspond
to many different equilibrium configurations, each very close to the
previous one. The slope of those steep parts of the curve is close to that of the initial, stiff rise
of $q$, confused with the axis on the main plot in the figure, and visible in the blown-up inset.
Large horizontal segments are due to motions between more distant configurations.
The origin of those two different regimes is clarified once it is attempted to find the system response to small load
increments by \emph{purely static} means. 
Starting from an equilibrium configuration, it is possible to regard its contact structure as
a given network of elastoplastic elements, and determine
the displacements leading to the new equilibrium configuration, with a static method which is a discrete analog of
elastoplastic finite element calculations in continuum mechanics. Such methods are seldom used
(see, however, \shortcite{KAK01}) in granular systems
because they are more complicated and less versatile than the usual dynamical approaches: a stiffness matrix
has to be rebuilt for each different contact list, and calculations are limited to the \emph{range of stability} of a
given contact network. As long as the contact structure is able to support the load, plastic strains
in the sliding contacts remain
contained by elastic strains in the non-sliding ones, and the static method is able to determine the sequence of
configurations reached on, \emph{e.g.,} stepwise increasing $q$. This sequence is made of a \emph{continuous set of
equilibrium states}, and the system evolution is indeed \emph{quasistatic}~: we refer to such
case as the \emph{strictly quasistatic r\'egime}.
We checked, for
series A samples, that static and dynamic calculations are in perfect agreement in such cases,
as shown on fig.~\ref{fig:dessmarfro}. This initial r\'egime is the stability range of
the initial configuration. The strains are  then directly due
to contact deformation -- such strains will be termed of \emph{type I}
in the sequel -- and are inversely proportional to $\kappa$,
while results are not sensitive to $\gamma$~ (the static method ignores completely inertia and physical time).
This range should
not be regarded as an elastic domain, as the non-linearity of the curves on fig.~\ref{fig:dessmarfro} 
(the elasticity of contacts is linear) is due to contact
losses and also to the gradual mobilization of friction. On reversing the $q$ increments, steeper slopes are observed. 
In the samples of fig.~\ref{fig:dessmarfro}, the very steep parts of the staircase-shaped curves also correspond, as we
checked, to stability intervals of some intermediate equilibrium configuration at higher $q$. Such intervals are separated
by large strain steps, corresponding to rearrangements of the contact structure. Those occur when the accumulation
of sliding contacts leads to an instability, and the ensuing motion is arrested by new contacts as interstices
between neighbouring grains are closed. The resulting strain increments are hereafter referred to
as \emph{type II} strains.
Their magnitude is related to the width of interstices between neighbouring grains.
The system evolution, in that
\emph{rearrangement r\'egime}, is, as shown previously, more sensitive to dynamical parameter
$\gamma$. Equilibrium states do not form a continuum in configuration space, the system has to jump between
two successive ones in a controlled deviator step test, or to flow nearby in a controlled strain rate test. The evolution
can only be termed quasistatic in a wider sense if the statistical properties of trajectories in configuration space are
independent, for slow enough motions, on dynamical parameters -- which can be reasonably expected from the present study.
The initial strictly quasi-static $q\le q_1$ interval does not shrink, but appears rather to approach a finite limit 
(about $q_1=0.8$ here) as the sample size increases. Stress-strain curves depend on $K_T/K_N$ within
this range, but, interestingly, $q_1$ does not~\shortcite{COMB2001}. 
In the rearrangement r\'egime, in order to approach a smooth curve in the
macroscopic limit (see fig.~\ref{fig:courbstat}), it is necessary that the sizes of both the steep and the flat parts
of the `staircases' shrink to zero as the sample size increases.
Type I and type II strains have very different amplitudes in A samples with $\kappa=10^5$ and $N\le 4900$. It might in fact
be expected that this clearcut distinction will get blurred at smaller stiffness parameter
$\kappa$ (whence larger type I strains) or larger $N$
(as smaller type II strain increments can close contacts), and that the transition at $q_1$ will be fuzzier.
Nevertheless, the system properties do strongly differ for $q<q_1$ and $q>q_1$, in two important respects. 
First, the slope of the
stress-strain curve relates directly to the elasticity
of the contacts in the type I strain dominated,
strictly quasistatic case. The tangent at the origin on
fig.~\ref{fig:dessmarfro} (smaller plot) is the Young modulus of the packing.  Second, 
the amplitude of fluctuations, the distance to
mechanical equilibrium, and the sensitivity to perturbations 
are much stronger in the rearrangement (type II strain dominated) r\'egime.
This is further illustrated by the following `creep experiment'~:
in a strain-rate controlled biaxial
compression, at some arbitrary instant, shift to stress-controlled conditions and keep $q$ constant, until an equilibrium
configuration is reached. Typical results of such tests are shown on fig.~\ref{fig:fluage}.
As could be expected, much larger strain variations are observed during periods of creep in the rearrangement regime, as
the initial states are farther from equilibrium. On resuming the constant strain rate test, 
the initial part of the curve is very steep, which is characteristic of a `strictly quasistatic' interval. From
an equilibrium state (devoid of sliding contacts), friction has to be mobilized again to produce the instabilities of the
rearrangement r\'egime. The dilatancy within those creep intervals is similar to the SRC one.

\begin{figure} 
\centering
 \includegraphics[width=9cm]{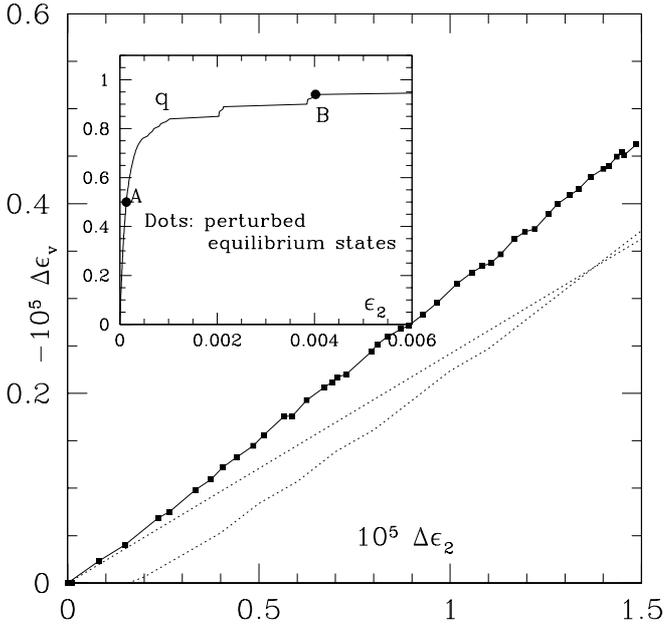}
\caption{
{\small
Effect of repeated random load ($f_0/aP = 2.5\, 10^{-3}$) applied in states shown as
big dots on the stress-strain curve in the inset: increments of $\epsilon_v$ vs. increments of $\epsilon_2$
on blown-up (by $10^5$) scale. The response of state A is concentrated near the origin, 
only the response of state B ($q=0.94$) is visible on this scale.
Dotted lines: SIC and SRC dilatancy curves near point B, same sample. }
\label{fig:pertu}
}
\end{figure}

The `creep tests' reveal different behaviours in
the two deformation regimes in SRC tests. One might also probe the sensitivity to perturbations of intermediate equilibrium
states obtained in SIC tests. We 
repeatedly applied on the grains constant external forces, each force
component being randomly chosen between $-f_0$ and $f_0$ ($f_0$ is a small fraction of $aP$), until new, perturbed equilibria
were reached.
Such random load increments always tend to produce strains in the same direction, as illustrated on fig.~\ref{fig:pertu}.
Applied when $q=0.5$ within the strictly quasistatic range, such perturbations
entail very small strain increments (hardly visible near the origin of the plot).
Applied when $q=0.94$ as equilibrium states are much
more unstable, they produce the series of strain increments plotted as connected dots,
which tend to accumulate proportionnally, hence the nearly straight line,
the slope of which is comparable to the dilatancy. The repeated application of small random perturbations
thus entails some `creep' phenomenon. 

\section{COMPARISONS WITH EXPERIMENTS}
In spite of the many differences between the numerical models and the materials studied in the laboratory, such as sand,
or even glass beads, some features of the simulation results can be compared in a qualitative or semi-quantitative
way to experimental ones. 

First, parameters $\kappa$ and $\gamma$ should be used to obtain robust estimations of orders of magnitude.
In 3 dimensions, $\kappa$ should be defined as 
$K_N /(aP)$ in the case of linear elasticity in the contacts. $\kappa$ measures the normal elastic deflection in a contact,
relatively to the grain diameter $a$, due to the typical 
contact force $Pa^2$. In a Hertzian contact between spheres of diameter $a$, it is easy to show that
$\kappa$ should be defined as $(E/P) ^{2/3}$, where $E$ is the Young modulus of the grain material. This gives
$\kappa \simeq 6000$ for glass beads under $P=10^5$Pa. (In 3D simulations, we could check that, given these definitions,
the effect of $\kappa$ on the coordination number was similar to the 2D case, see also~\shortcite{MJS00}).
`Real' materials with Hertz contacts under $P=10^5$Pa are
rather on the rigid side, but not quite in the rigid limit. Other contact laws
might lead to even smaller stiffness parameters (\emph{e.g.}, $\kappa \sim (E/P) ^{1/2}$ if $F_N\propto E h^2$,
as for cone-shaped asperities).

An appropriate 3D definition of $\gamma$ is $\dot \epsilon \sqrt{m\over aP}$ ($\sqrt{m\over aP}$ is the time for a grain
accelerated from rest by the typical force $a^2P$ to move on distance $a/2$). Substituting typical values -- a fraction
of millimetre for $a$, $10^{-5}s^{-1}$ for $\dot \epsilon$ -- this yields $\gamma$ values as small as $10^{-9}$ or
$10^{-10}$. As calculations over $\epsilon=2\%$ strain intervals with $\gamma = 10^{-5}$ still require several days
of c.p.u. time 
with 5000 stiff grains, real time scales of quasistatic laboratory tests are still beyond the reach of
discrete numerical simulations. $\gamma$ dependences of numerical results can however be extrapolated to smaller values.

Although it is tempting, in view of the
results illustrated on fig.~\ref{fig:fluage} to refer to creep experiments~\shortcite{MTK99,DiBT97},
as the aspects of the stress-strain curves
are quite similar in several respects, this difference of
time scales precludes a direct comparison. Moreover, the experimental $q$-$\epsilon$ curves do not depend on
strain rate if it is constant (this corresponds to much smaller $\gamma$ values than simulations), and
the creep deformation is extremely slow, often logarithmic in time~\cite{DiPI97}. Unlike in the numerical case, 
it does not appear to stop as some equilibrium is reached. It might well
be relevant, however, to discuss such experiments in terms of the sensitivity of the system to perturbations, 
which is likely to depend on whether contact networks resist load increments (strictly quasistatic case)
or are prone to instabilities (rearrangement r\'egime).
The numerical tests discussed in connection with fig.~\ref{fig:pertu}
suggest a possible microscopic origin of such slow evolutions over long times: a small noise level,
always present in experiments, could entail an accumulation of strain. 
Aging and creep phenomena can also be physically expected within one contact. Numerical simulations (devoid of
such features) might help assessing the collective aspects of the packing response.

Our simulations can also be likened to experimental observations
about the very small strain elastic behaviour of granular systems~\shortcite{diBGS99}.
Recent developments of precision apparati enabled measurements of strains in the
$10^{-5}$ range. To obtain elastic moduli, small stress cycles are superimposed on a constant loading,
producing cyclic strains on 
top of a systematic drift which, on increasing the number of cycles,
gradually slows down and becomes analogous to the one observed in creep tests.
The average slope of a cycle on
a stress-strain plot, once the effect of the drift is negligible, can be interpreted as an elastic modulus
(there remaining some small dissipation). Those small
strain increment elastic constants agree with the ones deduced from acoustic wave velocities. From our simulations, it 
transpires that the incremental stress-strain dependence might express a genuinely elastic behaviour (supplemented by some
plastic dissipation which vanishes in the limit of small stress increments) in the strictly quasistatic r\'egime. 
Elastic moduli are then related to the stiffness of the contacts. The width of strictly quasistatic strain intervals
are of the order of $q_1/\kappa$ -- $q_1$ being their width in terms of stress ratio. 
Taking into account that $q_1$ is exceptionally large for the initial small-strain r\'egime if our extremely dense and
well coordinated systems, and the value $\kappa \sim 6000$ estimated above for glass beads, one does obtain the
right order of magnitude ($\le 10^{-4}$) for the very small strain elastic domain. Moreover, the procedure by which these
moduli are measured can be interpreted as the preparation, either left to random perturbations or forced by
cyclic load increments, of a better stabilized state for which the contact network is able to resist small, but finite stress
increments (just like the stiffly responding equilibrium states of fig.~\ref{fig:fluage}).
\section{CONCLUSIONS AND PERSPECTIVES}
Despite their limitations
(due to the simplicity of the contact model, and the inaccessibility of long time scales),
the numerical simulation results presented here enable some investigation of the microscopic origins of
many features of experimentally observed behaviours. The definition of reduced dimensionles parameters ($\kappa$ and
$\gamma$) provides a framework in which many experimental and numerical studies can be discussed in common terms. 
Due to the small size of numerical samples, constitutive laws have to be approached via statistical analyses. Most
importantly, the distinction between two different origins of strain and two deformation
r\'egimes allows us some interpretations of
very small strain (tangential) elasticity and slow deformation (creep) under constant load, in terms of the system sensitivity to
perturbations. 

This work should be pursued in three directions. First, it is desirable to extend the existing approach to more `realistic' models,
so that more quantitative comparisons with experiments will be possible (our 3D results on spheres -- an obvious step in this
direction, were not presented here for lack of space). Secondly, the importance of the initial state and of the sample preparation
procedure calls for systematic studies (unlike for quasistatic monotonous compression tests, experimental knowledge is not 
expressed as well established laws for such processes). And, finally, the joint use of dynamic and static methods, which agree
remarkably in strictly quasistatic domains (fig.~\ref{fig:dessmarfro}) opens avenues to explore fundamental issues, such as
elastoplastic contact network stability and rearrangements, in some microscopic detail.

\begin{thebibliography}{}

{\small
\bibitem[\protect\citeauthoryear{Benahmed}{Benahmed}{2001}]{BENA1999}
Benahmed, N. (2001).
\newblock {\em Liqu\'efaction des sables}.
\newblock Ph.\ D. thesis, \'Ecole Nationale des Ponts et Chauss\'ees,
  Marne-la-Vall\'ee.

\bibitem[\protect\citeauthoryear{Combe}{Combe}{2001}]{COMB2001}
Combe, G. (2001).
\newblock {\em Origines g\'eom\'etrique du comportement quasi-statique des
  assemblages granulaires}.
\newblock Ph.\ D. thesis, \'Ecole Nationale des Ponts et Chauss\'ees,
  Marne-la-Vall\'ee.

\bibitem[\protect\citeauthoryear{Cundall and Strack}{Cundall and
  Strack}{1979}]{CUND79}
Cundall, P.~A. and O.~D.~L. Strack (1979).
\newblock A discrete numerical model for granular assemblies.
\newblock {\em G\'eotechnique\/}~{\em 29\/}(1), 47--65.

\bibitem[\protect\citeauthoryear{Di~Benedetto, Geoffroy, and
  Sauz\'eat}{Di~Benedetto et~al.}{1999}]{diBGS99}
Di~Benedetto, H., H.~Geoffroy, and C.~Sauz\'eat (1999).
\newblock Sand behaviour in very small to medium strain domains.
\newblock See \citeN{PDCG99}, pp.\  89--96.

\bibitem[\protect\citeauthoryear{Di~Benedetto and Tatsuoka}{Di~Benedetto and
  Tatsuoka}{1997}]{DiBT97}
Di~Benedetto, H. and F.~Tatsuoka (1997).
\newblock Small strain behaviour of geomaterials: modelling of strain rate
  effects.
\newblock {\em Soils and Foundations\/}~{\em 37\/}(2), 127--138.

\bibitem[\protect\citeauthoryear{Di~Prisco and Imposimato}{Di~Prisco and
  Imposimato}{1997}]{DiPI97}
Di~Prisco, C. and S.~Imposimato (1997).
\newblock Experimental analysis and theorical interpretation of triaxial load
  controlled loose sand specimen collapses.
\newblock {\em Mechanics of cohesive-frictional materials\/}~{\em 2}, 93--120.

\bibitem[\protect\citeauthoryear{Jamiolkowski}{Jamiolkowski}{1999}]{PDCG99}
Jamiolkowski, M. Lancellotta, R., and Lo~Presti, D. (Eds.) (1999).
\newblock {\em Pre-failure deformation characteristics of geomaterials},
  Rotterdam. Balkema.

\bibitem[\protect\citeauthoryear{Kishino}{Kishino}{2001}]{KI01}
Kishino, Y. (Ed.) (2001).
\newblock {\em Powders and Grains 2001}, Lisse. Swets \& Zeitlinger.

\bibitem[\protect\citeauthoryear{Kishino, Akaizawa, and Kaneko}{Kishino
  et~al.}{2001}]{KAK01}
Kishino, Y., H.~Akaizawa, and K.~Kaneko (2001).
\newblock On the plastic flow of granular materials.
\newblock See \citeN{KI01}, pp.\  199--203.

\bibitem[\protect\citeauthoryear{Kuhn}{Kuhn}{1999}]{KUHN99}
Kuhn, M.~R. (1999).
\newblock Structured deformation in granular materials.
\newblock {\em Mechanics of materials\/}~{\em 31}, 407--429.

\bibitem[\protect\citeauthoryear{Makse, Johnson, and Schwartz}{Makse
  et~al.}{2000}]{MJS00}
Makse, H., D.~Johnson, and L.~Schwartz (2000).
\newblock Packing of compressible granular materials.
\newblock {\em Physical Review Letters\/}~{\em 84\/}(18), 4160--4163.

\bibitem[\protect\citeauthoryear{Matsushita, Tatsuoka, Koseki, Cazacliu,
  Di~Benedetto, and Yasin}{Matsushita et~al.}{1999}]{MTK99}
Matsushita, M., F.~Tatsuoka, J.~Koseki, B.~Cazacliu, H.~Di~Benedetto, and
  S.~J.~M. Yasin (1999).
\newblock Time effects on the pre-peak deformation properties of sands.
\newblock See \citeN{PDCG99}, pp.\  681--689.

\bibitem[\protect\citeauthoryear{Roux}{Roux}{2000}]{JNR2000}
Roux, J.-N. (2000).
\newblock Geometric origin of mechanical properties of granular materials.
\newblock {\em Physical Review E\/}~{\em 61\/}(6), 6802--6836.

\bibitem[\protect\citeauthoryear{Roux and Combe}{Roux and Combe}{2002}]{RC02}
Roux, J.-N. and G.~Combe (2002).
\newblock Quasistatic rheology and the origins of strain.
\newblock {\em C. R. Acad\'emie des Sciences (Physique)\/}~{\em 3}, 131--140.

\bibitem[\protect\citeauthoryear{Silbert, Ertas, Grest, Halsey, and
  Levine}{Silbert et~al.}{2001}]{SEGHL01}
Silbert, L.~E., D.~Ertas, G.~S. Grest, T.~C. Halsey, and D.~Levine (2001).
\newblock Geometry of frictionless and frictional sphere packings.
\newblock {\em Phys.Rev. E\/}~{\em 64}, 051302.

\bibitem[\protect\citeauthoryear{Wood}{Wood}{1990}]{Muirwood}
Wood, D.~M. (1990).
\newblock {\em Soil Behaviour and Critical State Soil Mechanics}.
\newblock \,Cambridge University Press.

}
\end{thebibliography}

\end{document}